\documentclass{article} 
\usepackage[english]{babel}
\usepackage{amssymb}
\usepackage{latexsym}

\newenvironment{tab}{\begin{tabbing}
MMMMM\=aaa\=aaa\=aaa\=aaa\=aaa\=aaa\= \kill}{\end{tabbing}}

\makeatletter
\def\refmystepcounter#1{\stepcounter{#1}\protect\gdef 
\@currentlabel {\csname p@#1\endcsname \csname 
the#1\endcsname}}
\makeatother
\newcounter {tabnr}

\newcommand{\nrule}{\refmystepcounter{tabnr}
\mbreak\textbf{Lemma \arabic{tabnr}. }}
\def\la    {\mbox{$\langle$}}
\def\ra    {\mbox{$\rangle$}}
\def\boks    {\mbox{$\Box$}}
\def\bar   {\mbox{$[ \! ]$}}
\def\all   {\forall\;}
\def\ex    {\exists\;}
\def\sqleq {\sqsubseteq}
\def\S #1/{\mbox {\textsl{#1}}}
\def\B #1/{\mbox {\textbf{#1}}}
\def\R #1/{\mbox {\textrm{#1}}}
\def\T #1/{\mbox {\texttt{#1}}}
\def\Nat    {\mbox{$\mathbb{N}$}}
\def\sbreak {\smallbreak\noindent}
\def\mbreak {\medbreak\noindent}
\def\bbreak {\bigbreak\noindent}
\def\proof  {\mbreak{\sl Proof.\/} }
\def\IMPLIES{\mbox{$\quad\Rightarrow\quad $}}
\def\implies{\mbox{$\Rightarrow $}}
\def\LAND   {\mbox{\quad$\land$\quad}}
\def\Land   {\mbox{ $\;\land\;$ }}
\def\LOR    {\mbox{$\quad\lor\quad$}}
\def\Lor    {\mbox{ $\;\lor\;$ }}
\def\EQ     {\mbox{\quad$\equiv$\quad}}
\def\Eq     {\mbox{ $\;\equiv\;$ }}
\def\IS     {\mbox{$\quad =\quad $}}
\def\TO     {\mbox{$\quad\to\quad$}}
\def\true   {\S true/}
\def\false  {\S false/}
\def\DOTS   {\mbox{\ .\ .\ }}

\def\morph{\hbox{$\;-\!\triangleright\;$}}
\def\sem #1{\hbox{$[\![ \,#1\, ]\!] $}}
\def\sigma #1{{\hbox{{\sl states}($#1$)}}}
\def\alpha #1{{\hbox{{\sl start}($#1$)}}}
\def\nu #1{{\hbox{{\sl step}($#1$)}}}
\def\pi #1{{\hbox{{\sl prop}($#1$)}}}

\begin{document}
\setcounter{tabnr}{-1}
\begin {center}
{\Large\bf Eternity Variables to Prove
Simulation of Specifications}\\
\mbox{}\\
Wim H. Hesselink, \today\\
Dept. of Mathematics and Computing Science,
Rijksuniversiteit Groningen \\
P.O.Box 800, 9700 AV Groningen, 
The Netherlands\\
Email: \verb!wim@cs.rug.nl!, 
Web: {\tt http://www.cs.rug.nl/\~{}wim}\\
\end {center}

\begin{abstract}\noindent
Simulations of specifications are introduced 
as a unification and generalization of refinement 
mappings, history variables, forward simulations, 
prophecy variables, and backward simulations. 
A specification implements another specification if 
and only if there is a simulation from the 
first one to the second one that satisfies a certain 
condition. By adding stutterings, the formalism 
allows that the concrete behaviours take more (or
possibly less) steps than the abstract ones. 

Eternity variables are introduced as a more powerful 
alternative for prophecy variables and backward 
simulations. This formalism is semantically complete:
every simulation that preserves quiescence is a 
composition of a forward simulation, an extension with 
eternity variables, and a refinement mapping. This 
result does not need finite invisible nondeterminism and 
machine closure as in the Abadi-Lamport Theorem. 
The requirement of internal continuity is weakened to 
preservation of quiescence. 

Almost all concepts are illustrated by tiny examples or 
counter-examples. 
\end{abstract}

\bbreak {\bf Mathematics Subject Classification:}
68Q60, 68Q68.

\sbreak {\bf CR Categories:} F.1.1, F.3.1.

\sbreak \B Keywords / Implementations, simulations, 
history variables, prophecy variables, preservation of 
quiescence, refinement mapping, verification, 
invariants. 

\section{Introduction}

We propose eternity variables as a new formal tool to verify 
concurrent and distributed algorithms. 
Similar variables may have been used informally in the past 
in verifications as e.g.\ \cite{Bro92}. Eternity variables can 
also be applied to improve the abstractness and conciseness 
of specifications \cite {Hes03}. It is likely that they can 
be transferred to input-output automata, labelled transition 
systems, and perhaps even real-time and hybrid systems. 

Apart from proposing eternity variables and proving their 
soundness and completeness, this paper may serve as an 
introduction to the various forms of simulation for not 
necessarily terminating programs. We illustrate almost all 
concepts by tiny toy examples to sharpen the intuition. 

\subsection{Auxiliary Variables}

Eternity variables form a new kind of auxiliary variables,
variables that are added to a program to argue about it. 
Auxiliary variables occur when, in order to analyse a 
program, say $K$, one extends it with auxiliary variables 
and actions upon them to a bigger program, say $L$, proves 
some property of $L$, and infers something for the program, 
$K$, without them. 

Since the seventies, auxiliary variables have been used 
to prove the correctness of concurrent systems, e.g.\ 
\cite{Cli73,OwGr76}. These auxiliary variables served
to record the history of the system's behaviour. They are 
therefore sometimes called history variables. In e.g.\
\cite{Roe01}, it is proved that they are sufficient to 
prove that a terminating concurrent system satisfies a 
specification in terms of pre and postconditions. Such a 
result is called semantic completeness. 

In this paper, we want to allow nonterminating programs 
and therefore use ``abstract'' programs as specifications. 
The correctness issue then becomes the question of the 
implementation relation between programs. Over the years, 
the idea of implementation has been formalized in many 
different settings, under names like refinement and 
simulation. 

In or before 1986, it was proved that the combination of 
forward and backward simulations was sufficient to prove 
``data refinement'' for terminating programs \cite{HHS86}. 
In 1988, Abadi and Lamport \cite{AbL91} proposed prophecy 
variables to guess future behaviour of nonterminating 
programs. They proved that the combination of history 
variables, prophecy variables and refinement mappings is 
--in a certain sense-- sufficient to prove arbitrary 
implementation relations between nonterminating programs.
Although refinement mappings and extension with history 
variables can be regarded as forward simulations, and 
prophecy variables correspond to backward simulations, 
the two proofs of semantic completeness
are very different and the two papers \cite{AbL91,HHS86} 
do not refer to each other. They even have disjoint 
bibliographies.

The soundness of prophecy variables relies on K\"onig's 
Lemma; therefore, application of them requires that the 
invisible (i.e.\ internal) nondeterminism of the system 
is finite. One may argue that imposing finiteness 
should be acceptable since computer storage is always 
finite. Consider however the case that the prophecy would 
be the guess of a sequence number for the transactions in 
a reactive system, say an operating system or a database. 
Without a bound on the numbers, the choice would be 
infinite, but it is inacceptable to impose a bound on the 
number of transactions in the specification of such a 
system. Indeed, one would rather specify that the system 
can proceed indefinitely. In Sect.\ \ref{prophecy}, we 
give an example (H) to show the unsoundness of prophecy 
variables with a relation that allows infinite choices.

We therefore develop an alternative for prophecy 
variables that does not rely on K\"onig's Lemma. The 
eternity variable we propose as an alternative, is less 
flexible and it is chosen only once, nondeterministically 
and before the computation starts. 
Its value must of course be related to the behaviour as 
it develops. This will be dealt with in the so-called 
behaviour restriction. The proof of soundness for 
extension with eternity variables with a valid behaviour 
restriction is much easier than for prophecy variables. 

The new combination of extension with eternity variables 
and forward simulations is also proved to be semantically 
complete. This proof is somewhat easier than the 
corresponding proof for prophecy variables. We actually have 
two versions of this result, which differ in the degree of 
ignoring stutterings. 

\subsection{Additional Technical Assumptions}

Our setting is the theory of Abadi and Lamport \cite{AbL91},
where programs, systems, and specifications are all regarded 
as specifications. A specification 
is a state machine with a supplementary property. 
Behaviours of a specification are infinite sequences 
of states. Behaviours become visible by means of an 
observation function. A specification implements 
another one when all visible behaviours of the first 
one can occur as visible behaviours of the second one. 
Although they can change roles, let us call the 
implementing specification the concrete one and the 
implemented specification the abstract one. 

Under some technical assumptions, Abadi and Lamport 
\cite{AbL91} proved that, when 
a specification $K$ implements a specification $L$, 
there exists an extension $M$ of $K$ with history 
variables and prophecy variables together with a 
refinement mapping from $M$ to $L$. The assumptions 
needed are that $K$ should be ``machine closed'', and 
that $L$ should be ``internally continuous'' and of 
``finite invisible nondeterminism''. 

In our alternative with eternity variables instead of 
prophecy variables, ``internal continuity'' is weakened 
to ``preservation of quiescence'' while the other two 
assumptions are eliminated. Preservation of quiescence 
means that, whenever the concrete specification can 
repeat the current state indefinitely, the abstract 
specification is allowed to do so as well. In other 
words, when the implementation stops, the specification 
allows this. Preservation of quiescence is quite common. 
Indeed, refinement mappings and extensions with history, 
prophecy or eternity variables all preserve quiescence. 

\subsection{Stuttering Behaviour}

Since the concrete specification may have to perform 
computation steps that are not needed for the 
abstract specification, we follow \cite{AbL91,Lam94} 
by allowing all specifications to stutter: a behaviour 
remains a behaviour when a state in it is duplicated. 

In \cite{AbL91}, it is also allowed that the concrete 
specification is faster than the abstract one: 
a concrete behaviour may have to be slowed down by 
adding stutterings in order to match some abstract 
behaviour. This may seem questionable since one 
may argue that, when the concrete specification needs 
fewer steps than the abstract one, the abstract one is 
not abstract enough. Yet, experience shows that there 
need not be anything wrong with a specification when 
the implementation can do with fewer steps \cite{Lam89}.

We therefore developed two theories: a strict theory 
and a stuttering theory \cite{Hes04}. The stuttering 
theory corresponds to the setting of \cite{AbL91}, 
where the concrete specification can do both more and 
fewer steps than the abstract specification. 
In the strict theory, the concrete specification can 
do more but not fewer steps than the abstract 
specification. This results in a hierarchy of 
implementations that is finer than for the stuttering 
theory. In this paper we only present the strict 
theory, since it is simpler and more elegant than the 
stuttering theory of \cite{Hes04}.

\subsection{Simulations of Specifications}

A refinement mapping is a function between the states 
that, roughly speaking, preserves the initial states, 
the next-state relation and the supplementary 
property. Adding history or prophecy variables to the 
state gives rise to forward and backward simulations. 

We unify these three concepts by introducing simulations.
Actually, the term ``simulation'' has been introduced 
by Milner \cite{Mil71} in 1971. He used it for a kind 
of relation, which was later called downward or forward 
simulation to distinguish it from so-called upward 
or backward simulation \cite{HHS86,LyV95}. It seems 
natural and justified to reintroduce the term 
``simulation'' for the common generalization.

Our simulations are certain binary relations. 
For the sake of simplicity, we treat binary relations 
as sets of pairs, with some notational conventions.
Since we use $X\to Y$ for functions from $X$ to $Y$, 
and $P\:\implies\: Q$ for implication between predicates 
$P$ and $Q$, we write $F:K\morph L$ to denote that 
relation $F$ is a simulation of specifications from 
$K$ to $L$. We hope the reader is not confused by the 
totally unrelated arrows $\morph$ used in \cite{AbL95}.

The notation $F:K\morph L$ is inspired by category 
theory. Indeed, specifications with their simulations 
form the objects and morphisms of a category. Categories
were introduced in mathematics in \cite{EiM45}. Since 
every introduction to category theory goes far beyond  
our needs, we refrain from further references. 

Our first main result is a completeness theorem: a 
specification implements another one if and only if 
there is a certain simulation between them. 
This shows that our concept of simulation is 
general enough to capture the relevant phenomena. 

\subsection{Eternity Variables and Completeness}

In the field of program verification, simulations 
serve to prove correctness, i.e., the existence of an 
implementation relation between a program and a 
specification. The idea of refinement calculus is to 
construct simulations by composing them. Refinement 
mappings and forward simulations are the main 
candidates, but they are not enough. In general, one 
also needs simulations with kind of ``prescient 
behaviour'' as exhibited by backward simulations. It 
is at this point that our eternity variables come in. 

An eternity variable is a kind of logical variable
with a value constrained by the current execution. 
Technically, it is an auxiliary variable, which may be
initialized nondeterministically and is never modified
thereafter. Its value is constrained by a relation with 
the state. A behaviour that would violate such a 
constraint, is discarded. The verifier of a program 
has to prove that the totality of constraints is 
not contradictory. For example, the eternity variable
can be an infinite array while the conditions constrain 
different elements of it.

The simulation from the original specification to 
the one obtained by extending it with the eternity 
variable is called the eternity extension. 
We thus have four basic kinds of simulations: 
refinement mappings, forward simulations, backward 
simulations, and eternity extensions.
Every composition of simulations is a simulation. 
If relation $G$ contains a simulation $K\morph L$, then 
$G$ itself is a simulation $K\morph L$. Therefore, in 
order to prove that some relation $G$ is a simulation 
$K\morph L$, it suffices to find basic simulations such 
that the composition of them is contained in $G$. 
The completeness result is that, conversely, every 
simulation that preserves quiescence contains a 
composition of a forward simulation, an eternity 
extension, and a refinement mapping. 

More specifically, every specification $K$ has a 
so-called unfolding $K^\#$ \cite{LyV95} with a forward 
simulation $K\morph K^\#$. 
Given a simulation $F:K\morph L$ that preserves 
quiescence, we construct an intermediate 
specification $W$ as an extension of $K^\#$ with an 
eternity variable, together with a refinement 
mapping $W\morph L$, such that the composition of 
the simulations $K\morph K^\#$ and 
$K^\#\morph W$ and $W\morph L$ is a subset of relation
$F$. 

When one wants to use eternity variables to prove some
simulation relation, application of the unfolding $K^\#$
is overkill. Instead, one introduces approximating 
history variables to collect the relevant parts of the 
history. In Sect.\ \ref{behaveORassert}, we briefly 
discuss the methodological issues involved. A complete, 
but still tiny example is treated in Sect.\ \ref{example}. 
We refer to \cite{Hes03} for an actual application. 

\subsection{Overview}

In Sect.\ \ref{related}, we briefly discuss related 
work. Sect.\ \ref{tech} contains technical 
material on relations and lists. We treat stuttering and 
temporal operators in Sect.\ \ref{stut}. In Sect.\ 
\ref{formal}, we introduce specifications and simulations, 
and prove the characterizing theorem for them. In 
Sect.\ \ref{special}, we present the 
theory of forward and backward simulations in our 
setting and introduce quiescence and preservation of 
quiescence. Eternity variables are introduced in Sect.\
\ref{sec:refine}, where we also prove soundness and 
semantic completeness for eternity variables in the 
strict theory. Sect.\ \ref{example} contains a tiny 
application of the method: we consider a relation 
between the state spaces of two specifications and 
prove that it is a simulation by factoring it over a 
forward simulation, an eternity extension, an invariant 
restriction and two refinement mappings. 
Conclusions are drawn in Sect.\ \ref{conclusion}.
 
A preliminary version \cite{Hes02} of this paper 
was presented at MPC 2002. The paper \cite{Hes02} is 
flawed by an incorrect completeness theorem; we only 
saw the need of preservation of quiescence some weeks 
before the conference when the proceedings were already
in print. 

New concepts in this paper are simulation, preservation 
of quiescence, and eternity extension. 
New results are the completeness theorem of simulation 
with respect to implementation in Sect.\ \ref{visi}, the 
relationship between internal continuity and 
preservation of quiescence in Sect.\ \ref{strictsim}, 
and the soundness and completeness of eternity 
extensions in Sect.\ \ref{sec:refine}.

\subsection{Related Work}\label{related}

Our primary inspiration was \cite{AbL91} of Abadi and 
Lamport. Our formalism is a semantical version of 
Lamport's TLA \cite{Lam94}.
Lynch and Vaandrager \cite{LyV95} and Jonsson 
\cite{Jon91} present forward and backward simulations 
and the associated results on semantic completeness in 
the closely related settings of untimed automata and 
fair labelled transition systems. Our investigation 
was triggered by the paper \cite{CoL98} of Cohen and 
Lamport on Lipton's Theorem \cite{Lip75} about 
refining atomicity. While working on the serializable 
database interface problem of \cite {Lam92,Sch92}, we 
felt the need for variables with ``prescient'' behaviour 
without finiteness assumptions. This led us to the 
invention of eternity variables, which we applied 
successfully in the mean time to the serializable 
database interface in \cite{Hes03}. 
Jonsson, Pnueli, and Rump \cite{JPR99} present another 
way of proving refinement that avoids the 
finiteness assumptions of backward simulations. They 
use a very flexible concept of refinement based on 
so-called pomsets, but have no claim of semantic 
completeness.

\subsection{Relations and Lists}
\label{tech}

We treat a binary relation as a set of pairs. So, a 
binary relation between sets $X$ and $Y$ is a subset 
of the Cartesian product $X\times Y$. We use the 
functions \S fst/ and \S snd/ given by
$\S fst/(x,y) = x$ and $\S snd/(x,y) = y$.
A binary relation on $X$ is a subset of $X\times X$. 
The identity relation $1_X$ on $X$ consists of all 
pairs $(x,x)$ with $x\in X$. Recall that a binary 
relation $A$ on $X$ is called \emph{reflexive} iff 
$1_X\subseteq A$. The \emph{converse} $\S cv/(A)$ of 
a binary relation $A$ is defined by 
$\S cv/(A)=\{(x,y)\,|\,(y,x)\in A\}$. 

For binary relations $A$ and $B$, the composition 
$(A;B)$ is defined to consist of all pairs $(x,z)$ 
such that there exists $y$ with $(x,y)\in A$ 
and $(y,z)\in B$. A function $f:X\to Y$ is identified 
with its graph $\{(x,f(x))\,|\,x\in X\}$ which is a 
binary relation between $X$ and $Y$. The composition 
of functions $f:X\to Y$ and $g:Y\to Z$ is a function 
$g\circ f:X\to Z$, which equals the relational 
composition $(f;g)$. 

We use lists to represent consecutive values during 
computations. If $X$ is a set, we write $X^+$ for the 
set of the nonempty finite lists and $X^\omega$ for the 
set of infinite lists over $X$. 
We write $\ell(\S xs/)$ for the length of list 
\S xs/. The elements of \S xs/ are $\S xs/_i$ for 
$0\leq i < \ell(\S xs/)$.
If \S xs/ is a list of length $\ell(\S xs/)\geq n$, 
we define $(\S xs/\,|\,n)$ to be its prefix of length 
$n$. We write $\S xs/\sqleq \S xt/$ to denote that list 
\S xs/ is a prefix of \S xt/, possibly equal to \S xt/. 
We define $\S last/:X^+\to X$ to be the function that 
returns the last element of a nonempty finite list.

A function $f:X\to Y$ induces a function 
$f^\omega :X^\omega\to Y^\omega$.
For a binary relation $F\subseteq X\times Y$, we have 
an induced binary relation
$F^\omega\subseteq X^\omega\times Y^\omega$ given by
\begin{tab}
\> $ (\S xs/,\S ys/)\in F^\omega\EQ 
 (\all i::(\S xs/_i,\S ys/_i)\in F)$ .
\end{tab}

\subsection{Stuttering and Properties} \label{stut}

Let $P$ be a set of infinite lists over $X$, i.e., a 
subset of $X^\omega$. We write $\neg P$ 
to denote the complement (negation) of $P$. For an 
infinite list \S xs/, we write $\S Suf/(\S xs/)$ to 
denote the set of its infinite suffixes. The sets 
$\Box P$ (always $P$), and $\Diamond P$ (sometime $P$) 
are defined by
\begin{tab}
\> $ \S xs/\in\Box P\EQ \S Suf/(\S xs/)\subseteq P$ ,\\
\> $ \Diamond P \IS \neg\Box\neg P$ .
\end{tab}
So, $ \S xs/\in\Box P$ means that all suffixes of \S xs/ 
belong to $P$, and  $ \S xs/\in\Diamond P$ means that 
\S xs/ has some suffix that belongs to $P$.

For $U\subseteq X$ and $A\subseteq X\times X$, the 
subsets \sem {U} and \sem {A} of $X^\omega$ are 
defined by
\begin{tab}
\> $ \S xs/\in\sem {U}\EQ \S xs/_0\in U$ ,\\
\> $ \S xs/\in\sem {A}\EQ (\S xs/_0,\S xs/_1)\in A$ .
\end{tab}
So, $\sem U$ consists of the infinite lists that start 
in $U$, and $\sem A$ consists of the infinite lists that 
start with an $A$-transition.

We define a list \S xs/ to be an \emph {unstuttering} 
of a list \S ys/, notation $\S xs/\preceq\S ys/$, 
iff \S xs/ is obtained from \S ys/ by replacing some 
finite nonempty subsequences \S ss/ of consecutive 
equal elements of \S ys/ with their first elements 
$\S ss/_0$. The number of such subsequences that are
replaced may be infinite. For example, if, for a finite 
list \S vs/, we write $\S vs/^\omega$ to denote the list 
obtained by concatenating infinitely many copies of 
\S vs/, the list $(abbccb)^\omega$ is an unstuttering of 
$(aaabbbccb)^\omega$.

A finite list \S xs/ is called \emph{stutterfree} 
iff every pair of consecutive elements differ.
An infinite list \S xs/ is called \emph{stutterfree} 
iff it stutters only after reaching a final state, 
i.e., iff $\S xs/_i=\S xs/_{i+1}$ implies 
$\S xs/_{i+1}=\S xs/_{i+2}$ for all $i$. For every 
infinite list \S xs/, there is a unique stutterfree 
infinite list \S xt/ with $\S xt/ \preceq \S xs/$. For 
example, if $\S xs/ = (aaabbbccb)^\omega$ then 
$\S xt/ = (abcb)^\omega $.

A subset $P$ of $X^\omega$ is called a \emph{property
over} $X$ iff $\S xs/\preceq \S ys/$ implies that
$\S xs/\in P\Eq \S ys/\in P$. This definition is 
equivalent to the one of \cite{AbL91}. 
If $P$ is a property, then  $\neg P$, $\Box P$, and 
$\Diamond P$ are properties. If $U$ is a subset of $X$
then $\sem U$ is a property. If $A$ is a reflexive 
relation on $X$ then $\Box\sem A$ is a property, and it
consists of the infinite lists with all transitions 
belonging to $A$. 

\section{Specifications and Simulations}
\label{formal}

In this section we introduce the central concepts of 
the theory. Following \cite{AbL91}, we define 
specifications in Sect.\ \ref{specs}. Refinement mappings 
are introduced in Sect.\ \ref{refMap}. In \ref{synMor}, 
we define simulations. In Sect.\ \ref{visi} we define 
visible specifications and their implementation 
relations, and we prove that simulations characterize 
the implementations between visible specifications.

\subsection{Specifications} \label{specs}

A \emph{specification} is defined to be a tuple 
$K=(X,Y,N,P)$ 
where $X$ is a set, $Y$ is a subset of $X$, $N$ a 
reflexive binary relation on $X$, and $P$ is a 
property over $X$. The set $X$ is called the 
\emph{state space}, its elements are called 
\emph{states}, the elements of $Y$ are called 
\emph{initial states}. Relation $N$ is called the
\emph{next-state} relation. The set $P$ is called the 
\emph{supplementary} property.  

We define an \emph{initial execution} of $K$ to be a 
nonempty list \S xs/ over $X$ with $\S xs/_0\in Y$
and such that every pair of consecutive elements belongs 
to $N$. 
We define a \emph{behaviour} of $K$ to be an infinite 
initial execution \S xs/ of $K$ with $\S xs/\in P$. 
We write $\S Beh/(K)$ to denote the set of behaviours 
of $K$. 

The triple $(X, Y, N)$ can be regarded as a state machine
\cite{AbL91}. The supplementary property $P$ is often used 
for fairness conditions but can also be applied for other 
purposes. The initial executions of $K$ are determined by 
the state machine. The supplementary property is a 
restriction on the behaviours. 

It is easy to see that 
$\S Beh/(K) = \sem Y\cap \Box \sem N\cap P$. It follows 
that $\S Beh/(K)$ is a property.
The requirement that relation $N$ is reflexive is 
imposed to allow stuttering: if \S xs/ is a behaviour
of $K$, any list \S ys/ obtained from \S xs/ by 
repeating elements of \S xs/ or by removing 
subsequent duplicates is also a behaviour of $K$. 
In particular, for every behaviour \S xs/ of $K$, there 
is a unique stutterfree behaviour \S xt/ of $K$ with 
$\S xt/\preceq \S xs/$.

The components of specification $K = (X,Y,N,P)$
are denoted $\sigma K= X$, $\alpha K=Y$, 
$\nu K=N$ and $\pi K = P$. 

Specification $K$ is defined to be \emph{machine 
closed} \cite{AbL91} iff every finite initial execution 
of $K$ can be extended to a behaviour of $K$. We would 
encourage specifiers to write specifications that are 
not machine closed whenever that improves clarity, e.g., 
see \cite{LLOR99} Sect.\ 3.2.3.
If the specification is not machine closed, it is 
important to distinguish between states reachable from 
initial states and states that occur in behaviours. 

We therefore define a state of $K$ to be 
\emph{reachable} iff it occurs in an initial execution 
of $K$, and to be \emph{occurring} iff it occurs in a 
behaviour of $K$. A subset of $\sigma K$ is called a
\emph{forward invariant} iff it contains all reachable 
states. It is called an \emph{invariant} iff it contains 
all occurring states. Recall that a subset is called a 
\emph{strong invariant} (or inductive \cite{MaP95}) iff 
it contains all initial states and is preserved in every 
step, i.e. $J$ is a strong invariant iff $Y\subseteq J$ 
and $y\in J$ for every pair $(x,y)\in N$ with $x\in J$.  
It is easy to see that every strong invariant is a 
forward invariant and that every forward invariant is an 
invariant. 

\mbreak \S Example A./ 
Reachable states need not be occurring, an 
invariant need not be a forward invariant, and a forward 
invariant need not be a strong invariant. This is shown 
by the following program
\begin{tab}
\>\+ $ \B var /\;\T k/:\S Int/:= 0 $ ;\\
$ \B do /\;\T k/= 0 \TO \S choose /\;\T k/ > 0 $ ;\\
$\;\bar\quad \T k/\ne 0\TO \T k/:=\T k/-2 $ ;\\
\B od /;\\
\B prop: / infinitely often $\T k/=0$ .
\end{tab}
Note that this program only stands for a specification. 
It is not supposed to be directly executable. 

Formally, the specification is $(X,Y,N,P)$ where $X$ is 
the set of the integers and $Y=\{0\}$. A pair $(k,k')$ 
belongs to relation $N\subseteq X\times X$ if and only if 
\begin{tab}
\> $ (k=0\Land k' >0)\LOR
 (k\ne 0\Land k' = k-2)\LOR k' = k $ .
\end{tab}
The third disjunct serves to allow stuttering.
Property $P$ consists of the infinite sequences with 
infinitely many zeroes, i.e. $P=\Box\Diamond\sem Y$. It 
follows that the only occurring states are the 
even natural numbers. So, the even natural numbers form 
an invariant $J0$. The set of the natural numbers is 
also an invariant. The set of reachable states is 
$J1=\{k\,|\,k\geq 0 \Lor k\B\ mod /2 = 1\}$. Therefore 
$J0$ is not a forward invariant. The set $J1\cup\{-2\}$
is a forward invariant but not a strong invariant, since 
there is a step from $-2$ to $-4$. \boks

\subsection{Refinement Mappings}
\label{refMap}

Let $K$ and $L$ be specifications. 
A function $f:\sigma K\to \sigma L$ is called a 
\emph{refinement mapping} \cite{AbL91} from $K$ to 
$L$ iff $f(x)\in \alpha L$ for every $x\in \alpha K$, 
and $(f(x), f(x'))\in \nu L$ for every pair 
$(x,x')\in \nu K$, and $f^\omega (\S xs/)\in \pi L$ 
for every $\S xs/\in\S Beh/(K)$. 
Refinement mappings form the simplest way to compare 
different specifications.

\mbreak\S Example B./ For $m>1$, let $K(m)$ be the 
specification that corresponds to the program
\begin{tab}
\>\+ $ \B var /\;\T j/:\S Nat/:= 0 $ ;\\
$ \B do /\;\true \TO \T j/ := (\T j/+1)\B\ mod /m
\;\B\ od/$ ; \\
\B prop: / \T j/ changes infinitely often.
\end{tab}
We thus have $\sigma {K(m)}=\Nat$, $\alpha{K(m)} =\{0\}$,
$ \pi{K(m)} = \Box\Diamond\sem\ne $, and 
\begin{tab}
\>\+$(j,j')\in\nu{K(m)}\EQ j'\in\{j,(j+1)\B\ mod /m\}$ .
\end{tab}
In order to give an example of a refinement mapping, we
regard $K(20)$ as an implementation of $K(13)$. 
Let $f:\Nat\to\Nat$ be the function given by 
$ f(j) = \R min/(j,12)$. It is easy to 
verify that $f$ is a refinement mapping from $K(20)$ to 
$K(13)$. Note that the abstract behaviour (in $K(13)$) stutters 
whenever the concrete behaviour (in $K(20)$) is proceeding 
from $12$ to $19$. This example shows that it is useful 
that the next-state relation is always reflexive. \boks

\subsection{Simulations} \label{synMor}

Recall from \ref{tech} that a relation $F$ between 
\sigma K and \sigma L induces a relation $F^\omega$ 
between the sets of infinite lists $(\sigma K)^\omega$ 
and $(\sigma L)^\omega$.

We define relation $F$ to be a \emph{simulation} 
$K\morph L$ iff, for every behaviour 
$\S xs/\in \S Beh/(K)$, there exists a behaviour 
$\S ys/\in \S Beh/(L)$ with 
$(\S xs/,\S ys/)\in F^\omega$. 
The following two examples show that refinement 
mappings are not enough and that simulations are useful.

\mbreak\S Example C./ We use the specifications $K(m)$ 
and $K(2\cdot m)$ according to example B. Let the 
binary relation $F$ be given by
\begin{tab}
\> $ (j,k)\in F \EQ j=k\B\ mod /m $ .
\end{tab}
Then $F$ is a simulation $K(m)\morph K(2\cdot m)$, but 
there is no refinement mapping from $K(m)$ to
$K(2\cdot m)$. \boks

\mbreak\S Example D./ We consider two specifications
$K$ and $L$, both with state space $X=\{0,1,2,3,4\}$, 
initial set $Y=\{4\}$, and property 
$\Diamond\:\sem{\{0,1\}}$. The next-state relations are 
\begin{tab}
\> $ \nu K\IS 1_X \cup \{ (4,2), (2,1), (2,0)\} $ ,\\
\> $ \nu L\IS 1_X \cup \{(4,3), (4,2), (3,1), (2,0)\}$ .
\end{tab}

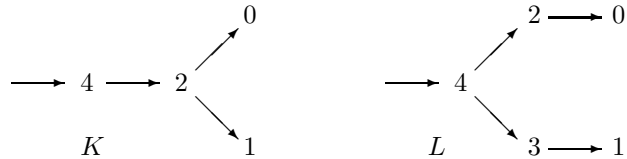
\setlength{\unitlength}{0.7mm}
\begin{picture}(120,35)(-16,-3)
\put(9,2)     {$K$}
\put(9,14)    {4}
\put(-4,15.5) {\vector(1,0){10}}
\put(14,15.5) {\vector(1,0){10}}
\put(27,14)   {2}
\put(31,18)   {\vector(1,1){8}}
\put(40,27)   {0}
\put(31,13)   {\vector(1,-1){8}}
\put(40,2)    {1}
\put(75,2)    {$L$}
\put(67,15.5) {\vector(1,0){10}}
\put(80,14)   {4}
\put(84,18)   {\vector(1,1){8}}
\put(94,27)   {2}
\put(84,13)   {\vector(1,-1){8}}
\put(94,2)    {3}
\put(98,28)   {\vector(1,0){10}}
\put(98,3)    {\vector(1,0){10}}
\put(110,27)  {0}
\put(110,2)   {1}
\end{picture}

\noindent
Both specifications have the final outcomes 0 and 1, but 
$K$ postpones the choice, while $L$ chooses immediately.
We regard only the final states 0 and 1 as visible. The 
stutterfree behaviours of $K$ are $(4,2,0^\omega)$ 
and $(4,2,1^\omega)$, while those of $L$ are  
$(4,2,0^\omega)$ and $(4,3,1^\omega)$.
Therefore, $K$ and $L$ implement each other. One can 
easily verify that relation $F=1_X\cup\{(2,3)\}$ is a 
simulation $F:K\morph L$. There is no refinement mapping 
$f$ from $K$ to $L$ with $f(0) = 0$ and $f(1) = 1$, 
since the concrete specification $K$ makes the 
choice between the outcomes later than the abstract 
specification $L$. At concrete state 2, simulation $F$ 
``needs prescience'' to choose between the abstract 
states 2 and 3. \boks

\bigbreak In general, it should be noted that the mere 
existence of a simulation $F:K\morph L$ does not imply 
much. If $F:K\morph L$ and $G$ is a relation with 
$F\subseteq G$, then $G:K\morph L$. Therefore, the 
smaller the simulation, the more information it 
carries. 
It is easy to verify that simulations can be 
composed: if $F$ is a simulation $K\morph L$
and $G$ is a simulation $L\morph M$, the 
composed relation $(F;G)$ is a simulation 
$K\morph M$. 
It is also easy to verify that a refinement mapping 
$f:\sigma K\to \sigma L$, when regarded as a relation 
as in Sect.\ \ref{tech}, is a simulation $K\morph L$. 

We often encounter the following situation. A 
specification $L$ is regarded as an {extension} of 
specification $K$ with a variable of a type $M$ iff 
$\sigma L$ is (a subset of) the Cartesian product 
$\sigma K\times M$ and the function 
$\S fst/:\sigma L\to \sigma K$ is a refinement mapping.
The second component of the states of $L$ is then 
regarded as the variable added. The extension is
called a {refinement extension} iff the converse
$\S cv/(\S fst/)$ is a simulation $K\morph L$.

\subsection{Visibility and Completeness of Simulation}
\label{visi}

We are usually not interested in all details of the
states, but only in certain aspects of them. This 
means that there is a function from $\sigma K$ to 
some other set that we regard as an observation 
function. A \emph{visible specification} 
is therefore defined to be a pair $(K,f)$ where 
$K$ is a specification and $f$ is some function 
defined on $\sigma K$. 
Deviating from \cite{AbL91}, we define the set of 
observations by
\begin{tab}
\> $\S Obs/(K,f)\IS \{f^\omega (\S xs/) \,
|\, \S xs/\in \S Beh/(K)\}$ .
\end{tab}
Note that $\S Obs/(K,f)$ need not be a property. If
\S xs/ is an observation and $\S ys/\preceq\S xs/$, 
then \S ys/ need not be an observation.  

\mbreak\S Example E./ Assume we are observing $K(13)$ 
of example B with the test $j >0$. So, we use the 
observation function $f(j) = (j> 0)$. Then the 
observations are the boolean lists with infinitely 
many values \true\ and infinitely many values \false, 
in which every \true\ stutters at least 12 times.
\boks

\medbreak
Let $(K,f)$ and $(L,g)$ be visible specifications
with the functions $f$ and $g$ mapping to the same 
set. Then $(K,f)$ is said to \emph{implement}  
$(L,g)$ iff $\S Obs/(K,f)$ is contained in 
$\S Obs/(L,g)$, i.e., iff for every $\S xs/ \in\S Beh/(K)$
there exists  $\S ys/ \in\S Beh/(L)$ with 
$f^\omega(\S xs/) = g^\omega(\S ys/)$. 
This concept of implementation is stronger than 
that of \cite{AbL91}: we do not allow that an 
observation of $(K,f)$ can only be mimicked by $(L,g)$
after inserting additional stutterings. 

Our concept of simulation is motivated by the
following completeness theorem, the proof of
which is rather straightforward.

\bbreak
\B Theorem 0./ Consider visible specifications $(K,f)$ 
and $(L,g)$ where $f$ and $g$ are functions to 
the same set. We have that $(K,f)$ implements $(L,g)$ 
if and only if there is a simulation $F:K\morph L$ 
with $(F;g)\subseteq f$. 

\proof The proof is by mutual implication. 

First, assume the existence of a simulation
$F:K\morph L$ with $(F;g)\subseteq f$. 
Let $\S zs/\in\S Obs/(K,f)$. We have to prove 
that $\S zs/\in \S Obs/(L,g)$. By the definition of 
\S Obs/, there exists $\S xs/\in\S Beh/(K)$ with 
$\S zs/ = f^\omega(\S xs/)$. Since $F$ is a 
simulation, there exists $\S ys/\in \S Beh/(L)$ with 
$(\S xs/, \S ys/)\in F^\omega$. For every number $n$, 
we have $(\S xs/_n,\S ys/_n)\in F$ and, hence, 
$(\S xs/_n,g(\S ys/_n))\in (F;g)\subseteq f$ and, 
hence, $g(\S ys/_n)=f(\S xs/_n)= \S zs/_n$. This 
implies that 
$\S zs/ = g^\omega(\S ys/) \in \S Obs/(L,g)$.

Next, assume that $(K,f)$ implements $(L,g)$.
We define relation $F$ between $\sigma K$ and 
$\sigma L$ by $F=\{(x,y)\,|\,f(x) = g(y)\}$. For 
every pair $(x,z)\in (F;g)$ there exists $y$ with 
$(x,y)\in F$ and $(y,z)\in g$; we then have 
$f(x)=g(y)=z$. This proves $(F;g)\subseteq f$. 
It remains to prove that $F$ is a simulation 
$K\morph L$. Let $\S xs/\in\S Beh/(K)$. Since 
$\S Obs/(K,f)\subseteq \S Obs/(L,g)$, there is 
$\S ys/\in\S Beh/(L)$ with 
$f^\omega(\S xs/) = g^\omega(\S ys/)$. We thus have
$(\S xs/,\S ys/)\in F^\omega$. This 
proves that $F$ is a simulation $K\morph L$. \boks

\mbreak\S Example F./ Consider the visible 
specifications $(K,f)$ and $(L,g)$ with $K=K(m)$ 
and $L=K(2\cdot m)$ as in example C, with $f$, 
$g:\Nat\to\Nat$ given by $f(j)=j$ and $g(j)=j\B\ mod /m$.
Then relation $F$ as constructed in the above proof 
equals relation $F$ of example C. \boks

\section{Special Simulations}\label{special}

In this section we introduce forward and backward 
simulations as special kinds of simulations. 
Forward simulations are introduced in \ref{forward}. 
They correspond to refinement mappings 
and to the well-known addition of history variables.
In \ref{invarRes}, we show that invariants give rise to  
simulations.
In Sect.\ \ref{histComp}, we introduce the unfolding 
\cite{LyV95} of a specification, which plays a key 
role in several proofs of semantic completeness. 
Backward simulations are introduced in Sect.\ 
\ref{prophecy}. Quiescence and preservation of 
quiescence are introduced in Sect.\ \ref{strictsim}.

\subsection{Flatness and Forward Simulations}\label{forward}

We start with a technical definition concerning the 
supplementary property of the related specifications. 
A relation $F$ between \sigma K and \sigma L is 
defined to be \emph{flat from $K$ to $L$} iff every 
infinite initial execution \S ys/ of $L$ with 
$(\S xs/,\S ys/) \in F^\omega$ for some 
$\S xs/ \in \S Beh/(K)$ satisfies $\S ys/\in\pi L$. 

It turns out that all our basic kinds of simulations 
are flat. Indeed, refinement mappings are flat and we 
need flatness as a defining condition for both 
forward and backward simulations. Flatness always 
serves as the finishing touch in the construction of 
the abstract behaviour. Yet, flatness is 
not a nice property: in example G below, we show 
that the composition of two flat simulations need 
not be flat. 

The easiest way to prove that one specification 
simulates (the behaviour of) another is by starting 
at the beginning and constructing the corresponding 
behaviour in the other specification inductively. 
This requires a condition embodied in so-called 
forward or downward simulations \cite{HHS86,LyV95}, 
which go back at least to \cite{Mil71}. They are 
defined as follows. 

A relation $F$ between \sigma K and \sigma L is 
defined to be a \emph{forward simulation} from 
specification $K$ to specification $L$ iff 

\mbreak
(F0) \ For every $x\in\alpha K$, there is 
$y\in\alpha L$ with $(x,y)\in F$.\\
(F1) \ For every pair $(x,y)\in F$ and every $x'$ 
with $(x,x')\in \nu K$, there is $y'$ with 
$(y,y')\in\nu L$ and $(x',y')\in F$.\\
(F2) \ Relation $F$ is flat from $K$ to $L$. 

\mbreak\S Examples./ It is easy to verify that relation 
$F$ of example C is a forward simulation.
Every refinement mapping, when regarded as a relation, 
is a forward simulation. \boks

\smallbreak
The definition of forward simulations is 
justified by the following well-known result:

\mbreak
\B Lemma/. Every forward simulation $F$ from $K$ to 
$L$ is a simulation $K\morph L$. 

\proof Let $\S xs/\in \S Beh/(K)$ be given. Then 
$\S xs/_0\in \alpha K$, so by (F0), there is 
$\S ys/_0\in\alpha L $ with 
$(\S xs/_0,\S ys/_0)\in F$. 
Since $(\S xs/_n,\S xs/_{n+1})\in \nu K$ for all 
$n$, we can use (F1) inductively to construct an 
infinite initial execution \S ys/ of $L$ that 
satisfies $(\S xs/_n,\S ys/_n)\in F$ for all $n$. 
Since relation $F$ is flat, we conclude that \S ys/ 
is a behaviour of $L$ with 
$(\S xs/,\S ys/)\in F^\omega$. Therefore $F$ 
is a simulation $K\morph L$. \boks

\mbreak\S Example G./ Let $X=[0\DOTS N]$ for some number 
$N\geq 2$. Let $K$ be the specification with the program 
\begin{tab}
\>\+ $ \B var /\;\T k/:X:= 0 $ ;\\
$\B do /\;\true \TO \S choose / \T k/\in X\quad \B od/ $ ;\\
\B prop: /\T k/ changes infinitely often and is sometimes 1.
\end{tab}
So, we have $\sigma K = X$, $\alpha K = \{0\}$, and 
$\nu K = X^2$. The property $\pi K$ is the intersection of 
$\Box\Diamond\sem\ne$ and $\Diamond\,\sem{\T k/=1}$.

Let $L$ be the specification with 
\begin{tab}
\>\+ $ \B var /\;\T j/:X:= 0 $ ,\\
\> $ \T b/:\S Boolean/:=\false $ ;\\
$\B do /\;\true \TO \S choose / \T j/\in X $ ;\\
$\bar\quad \T j/ = 1 \TO \T b/:=\true\;;\;
\S choose / \T j/\in X$ ;\\
$ \B od/ $ ;\\
\B prop: /\T b/ is sometimes \true.
\end{tab}
In such programs, we regard the alternatives in de \B do/ 
loop as atomic. So we have 
\begin{tab}
\> $ ((j,b),(j',b'))\in\nu L \EQ (b'=b)\Lor (j=1\Land b') $ .
\end{tab}
The property is $\pi L = \Diamond\,\sem{\T b/}$.

It is easy to show that relation $F=\{(k,(j,b))\,|\,k=j\}$ 
is a simulation $K\morph L$. Indeed, let \S xs/ be a 
behaviour of $K$. Then there is an index $r$ with 
$\S xs/_r=1$. Let \S ys/ be the sequence in $\sigma L$ 
given by $\S ys/_i=(\S xs/_i, (r<i))$ for all $i$. Since 
the boolean component \T b/ of \S ys/ becomes true in a 
step with precondition $\T j/=1$, this is a behaviour of 
$L$, which satisfies $(\S xs/,\S ys/)\in F^\omega$. 
Simulation $F:K\morph L$ is not flat, since the sequence 
\S zs/ with $\S zs/_i=(\S xs/_i,\false)$ for all $i$ is 
not a behaviour of $L$ but is an infinite initial 
execution of $L$ with $(\S xs/,\S zs/)\in F^\omega$.

In order to show that $F$ is a composition of two forward 
simulations, we make specification $L$ more deterministic.
Let $L'$ be the specification obtained from $L$ by 
restricting the step relation to 
\begin{tab}
\>\+ $\B do /\; \T j/ \ne 1 \TO 
\S choose / \T j/\in X $ ;\\
$\bar\quad \T j/ = 1 \TO \T b/:=\true\;;\;
\S choose / \T j/\in X$ ;\\
$ \B od/ $ .
\end{tab}
Since stuttering must be allowed, a pair $((j,b),(j',b'))$ 
belongs to $ \nu {L'}$ if and only if
\begin{tab}
\> $ b'=(b\Lor j=1)\LOR (j=j'\Land b=b')$ . 
\end{tab}
The above relation $F$ is a forward simulation 
$K\morph L'$. Indeed, condition (F0) is obvious.
Condition (F1) holds since every step of $K$ can be 
mimicked by $L'$. Flatness is shown as follows. Let \S xs/ 
be a behaviour of $K$. The property of $K$ implies that 
there is an index $r$ with $\S xs/_r=1\ne \S xs/_{r+1}$. 
If \S ys/ is an infinite initial execution of $L'$ with 
$(\S xs/,\S ys/)\in F^\omega$, then \S ys/ is a behaviour 
of $L'$ since $\S ys/_{r+1}=(\S xs/_{r+1}, \true)$.

It is easy to verify that the identity function \S id/ 
is a refinement mapping $L'\morph L$ and hence a forward 
simulation. The simulation $F:K\morph L$ is clearly the 
composition $F=(F;\S id/)$. So, here we have indeed a 
nonflat composition of two forward simulations. \boks

\subsection{Invariant Restriction}\label {invarRes}

Invariants are often used to restrict the state space 
implicitly. When the state space is made explicit, 
restriction to an invariant subspace turns out to be 
a simulation. 

Slightly more general, let $D$ be a subset of 
$\sigma K$ for a specification $K$. Then we can define 
the $D$-restricted specification $K_D$ by 
$\sigma{K_D}= D$ and $\alpha{K_D}= D\cap \alpha K$ and 
$\nu{K_D}= D^2\cap \nu K$ and 
$\pi{K_D}= D^\omega\cap \pi K$. Indeed, it is easy to 
verify that $\nu{K_D}$ is reflexive and that $\pi{K_D}$
is a property. The following result characterizes 
invariants via simulations. 

\nrule\label{invars} (a) The identity relation 
$1_D$ is a simulation $K\morph K_D$ 
if and only if $D$ is an invariant.\\
(b) $1_D$ is a forward simulation $K\morph K_D$ if and 
only if $D$ is a strong invariant. 

\mbreak We skip the proof, since it is fairly 
straightforward and not interesting.

\subsection{The Unfolding}\label{histComp}

The \emph{unfolding} $ K^\# $ of a specification $K$
plays a key role in the proofs of semantic 
completeness in \cite{AbL91,LyV95} as well as in our
semantic completeness result below.

It is defined as follows: 
$\sigma {K^\#}$ consists of the stutterfree 
finite initial executions of $K$. The initial 
set $\alpha {K^\#}$ consists of the elements 
$\S xs/\in\sigma {K^\#}$ with $\ell(\S xs/)=1$. 
The next-state relation $\nu {K^\#}$ and the property 
$\pi {K^\#}\subseteq (\sigma {K^\#})^\omega$ are 
defined by 
\begin{tab}
\> $ (\S xs/,\S xt/)\in \nu {K^\#} \EQ
\S xs/ \sqleq \S xt/\Land 
\ell(\S xt/) \leq \ell(\S xs/) + 1$ ,\\
\> $ \S vss/\in \pi {K^\#} \EQ 
\S last/^\omega(\S vss/)\in \pi K$ .
\end{tab}
So, the nonstuttering steps of $K^\#$ are the pairs 
 $ (\S xs/,\S xt/)$ with $\S xs/ \sqleq \S xt/$ and
$\ell(\S xt/) = \ell(\S xs/) + 1$.

It is easy to prove that $K^\#$ is a specification. 
The function $ \S last/:\sigma {K^\#} \to \sigma K$ is 
a refinement mapping. Moreover, if 
$(\S xs/,\S xt/)\in\nu {K^\#}$ and $\S xs/\ne\S xt/$, 
then $\S last/(\S xs/)\ne\S last/(\S xt/)$ since \S xt/ 
is stutterfree. We are more interested, however, in 
the other direction. The following result of 
\cite{AbL91} is not difficult to prove.

\nrule\label{unfold}
Relation $ \S cvl/ = \S cv/(\S last/)$ is 
a forward simulation $ K\morph K^\#$. \boks

\bigbreak
In Sect.\ \ref{eterComp} below, we shall need the 
following result.

\nrule\label{bloweddown}
Let $ \S xs/ = \S last/^\omega(\S vss/) $ for a 
stutterfree behaviour \S vss/ of $ K^\# $. 
Then \S xs/ is a behaviour of $ K $ with 
$\: \S vss/_i\sqleq \S xs/\: $ for all indices $i$. 

\proof Since \S vss/ is a behaviour of $K^\#$, it is 
easy to verify that \S xs/ is a behaviour of $K$. 
We now distinguish two cases. First, assume that 
$\S vss/_i\ne \S vss/_{i+1}$ for all $i$. Then 
$\ell(\S vss/_i) = i+1$ for all $i$. It follows that 
$\S vss/_i = (\S xs/\,|\,i+1)$ for all $i$. Otherwise, 
let $r$ be minimal with $\S vss/_r = \S vss/_{r+1}$. 
Since \S vss/ is stutterfree, $\S vss/_i = \S vss/_r $ for 
all $i\geq r$. This implies 
$\ell(\S vss/_i) = \R min/(i,r)+1 $ for all indices $i$.
It follows that $\S vss/_i = (\S xs/\,|\,i+1)$ for all $i$ 
with $0\leq i \le r$ and $\S vss/_i=(\S xs/\,|\,r+1)$ for 
all $i$ with $r\leq i < \infty$. In either case, we have 
$\: \S vss/_i\sqleq \S xs/\: $ for all indices $i$. \boks

\subsection{Backward Simulations}\label{prophecy}

It is also possible to prove that one specification 
simulates (the behaviour of) another by starting 
arbitrarily far in the future and constructing a
corresponding initial execution by working backwards. 
An infinite behaviour is then obtained by a variation 
of K\"onig's Lemma. These so-called backward simulations 
\cite{LyV95} form a relational version of the prophecy 
variables of \cite{AbL91} and are related to the upward 
simulations of \cite{HHS86}.
We give a variation of Jonnson's version \cite{Jon91}. 

Relation $F$ between \sigma K and \sigma L is defined 
to be a \emph{backward simulation} from $K$ to $L$ iff 

\sbreak
(B0) \ Every pair $(x,y) \in F$ with $x\in \alpha K$ 
satisfies $y\in \alpha L$.\\
(B1) \ For every pair $(x',y')\in F$ and every $x$ 
with $(x,x')\in \nu K$, there is $y$ with $(x,y)\in F$ 
and $(y,y')\in\nu L$.\\
(B2) \ For every behaviour \S xs/ of $K$ there are 
infinitely many indices $n$ for which the set 
$\{y\mid (\S xs/_n,y)\in F\}$ is nonempty and finite.\\
(B3) \ Relation $F$ is flat from $K$ to $L$. 

\smallbreak The simulation $F$ presented in the example 
D in \ref{synMor} is a very simple example of a backward 
simulation. The verification of this is straightforward, 
though somewhat cumbersome. 

An auxiliary variable added to the state 
space via a backward simulation is called a prophecy 
variable \cite{AbL91} since it seems to show 
``prescient'' behaviour. In such a case, the relation 
is called a prophecy relation in \cite{LyV95}.  
The term backward simulations is justified by the 
following soundness result, the proof of which is a 
direct adaptation of the proof in \cite{Jon91}.

\mbreak
\B Lemma/. Every backward simulation $F$ from $K$ to 
$L$ is a simulation $K\morph L$. \boks

\medbreak The empty relation $F=\emptyset$ always satisfies 
(B0), (B1), and (B3), but if $K$ has any behaviour, the 
empty relation is not a simulation from $K$ to $L$. This 
justifies the nonemptyness condition in (B2). The following 
example shows that some finiteness in (B2) is also needed. 

\mbreak \S Example H: the unsound doomsday prophet./ 
Let $L$ be the following extension of specification $K(13)$
of example B with a natural variable \T k/. 
\begin{tab}
\>\+ $ \B var /\;\T j/:\S Nat/:= 0\;,\;\T k/:\S Nat/\; 
\{\S arbitrary/\} $ ;\\
$ \B do /\;\T k/> 0 \TO \T j/ := (\T j/+1)\B\ mod /13\;;\;
\T k/:= \T k/-1 \;\B\ od/$ ; \\
\B prop: / \T j/ changes infinitely often.
\end{tab}
Since \T k/ cannot decrease infinitely often, \T j/ cannot 
change infinitely often. Therefore, specification $L$ has no 
behaviours. Since $K(13)$ has behaviours, there cannot exist 
any simulation $K(13)\morph L$. Function 
$\S fst/:\sigma L\to\sigma{K(13)} $ is a refinement mapping.
Its converse, $F=\S cv/(\S fst/)$ cannot be a simulation 
$K(13)\morph L$, but it is easily seen to satisfy (B0), 
(B1), and (B3). Indeed, it does not satisfy (B2) since 
$\{y\mid (x,y)\in F\}$ is infinite for every 
$x\in \sigma{K(13)}$. 

The initial value of \T k/ can be regarded as a prophecy 
of doomsday, whence the name of the example. Note that 
$K(13)$ is deterministic and that the only nondeterminism 
in $L$ is the infinite choice in the initialization. 
Also, note that we can restore condition (B2) by introducing 
a bound, say $\T k/ < 1000$, for the initial choice of 
\T k/, but then condition (B1) is invalidated. \boks

\subsection{Preservation of Quiescence}\label{strictsim}

The completeness result of the next section needs the concept
of ``preservation of quiescence''. Roughly speaking, a 
behaviour is quiescent at a given state if it remains a 
behaviour when the behaviour after the state is replaced by 
an infinite repetition of the state. Preservation of 
quiescence means that the abstract behaviour can be quiescent 
whenever the concrete behaviour is quiescent. It is 
formalized as follows.
 
Given a natural number $n$ and an infinite list \S xs/, 
we define the infinite list 
$E_n(\S xs/)$ as the concatenation of $(\S xs/|n)$ 
with the infinite repetition of the state $\S xs/_n$. We 
thus have  $ (E_n(\S xs/))_k = \S xs/_m$ where $m$ is the 
minimum of $n$ and $k$. A number $n$ is a quiescent index 
of \S xs/ for specification $K$ iff $E_n(\S xs/)$ is a 
behaviour of $K$. The set of quiescent indices of \S xs/ 
for $K$ is defined as
\begin{tab}
\> $Q_K(\S xs/) = \{n\mid E_n(\S xs/)\in \S Beh/(K)\}$ . 
\end{tab}
Let $K$ and $L$ be specifications. A simulation 
$F:K\morph L$ is said to \emph{preserve quiescence} 
iff, for every $\S xs/\in\S Beh/(K)$, there exists 
$\S ys/\in \S Beh/(L)$ with 
$(\S xs/, \S ys/)\in F^\omega$ and 
$ Q_K(\S xs/)\subseteq Q_L(\S ys/)$. 

It is easy to verify that preservation of quiescence is 
compositional: if $F:K\morph L$ and 
$G:L\morph M$ both preserve quiescence, the composition 
$(F;G): K\morph M$ also preserves quiescence. Also, if 
$F:K\morph L$ preserves quiescence and $G$ is a relation 
between $\sigma K$ and $\sigma L$ with $F\subseteq G$, 
then $G$ is a simulation $K\morph L$ that preserves 
quiescence.

\mbreak\S Example G'./ Going back to example G in 
Sect.\ \ref{forward}, we let $K'$ be the specification 
obtained from $K$ by omitting the requirement that \T k/ 
keeps changing. So, the property is weakened to 
$\pi {K'} = \Diamond\sem {\T k/=1}$. By the 
same argument as before, relation $F$ is a simulation 
$K'\morph L'$. This simulation does not preserve 
quiescence. Indeed, let \S xs/ and \S ys/ be behaviours of 
$K'$ and $L'$ with $(\S xs/, \S ys/)\in F^\omega$. Let
$r$ is the first index with $\S xs/_r=1$, then $r$ is a 
quiescent index of \S xs/ but not of \S ys/, since
the boolean \T b/ is still false. \boks

\mbreak\S Example I./ We construct an even simpler 
simulation that does not preserve quiescence. Consider 
specifications $K$ and $L$, both with state space 
$X = \{0,1,2\}$, initial set $\{1\}$, and supplementary 
property $\Diamond\Box\:\sem{\{0\}}$. 
The next-state relations are given by 
\begin{tab}
\>\+ $\nu {K}\IS 1_X\cup\{(1,0), (0,1)\}$ ,\\
$\nu {L}\IS 1_X\cup\{(1,0), (1,2), (2,1)\}$ .
\end{tab}

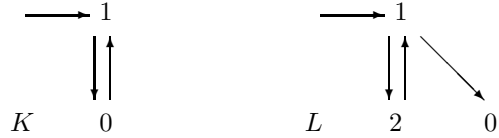
\setlength{\unitlength}{0.7mm}
\begin{picture}(120,35)(-20,0)
\put(0,29)    {\vector(1,0){12}}
\put(14,28)   {1}
\put(13,25)   {\vector(0,-1){12}}
\put(16,13)   {\vector(0,1){12}}
\put(14,7)    {0}
\put(-3,7)    {$K$}
\put(53,7)    {$L$}
\put(56,29)   {\vector(1,0){12}}
\put(70,28)   {1}
\put(69,25)   {\vector(0,-1){12}}
\put(75,25)   {\vector(1,-1){12}}
\put(72,13)   {\vector(0,1){12}}
\put(69,7)    {2}
\put(87,7)    {0}
\end{picture}

\noindent
The behaviours of $K$ are infinite lists over $\{0,1\}$ 
that start with 1 and contain only finitely many 
ones. The behaviours of $L$ are finite lists over 
$\{1,2\}$ that start and end with 1, followed by 
infinitely many zeroes. In either case, the quiescent 
indices are those of the zero elements in the list. 

Let relation $F$ on $X$ be the set 
$ F=\{(0,0), (0,2), (1,1)\}$.
Relation $F$ is a simulation $K\morph L$. In fact, 
for every $ \S xs/ \in\S Beh/(K)$, there is precisely one
$ \S ys/ \in\S Beh/(L)$ with 
$(\S xs/,\S ys/)\in F^\omega$. If $n$ is the least 
number with $\S xs/_i = 0$ for all $i\geq n$, then 
$\S ys/_j=2$ for all $j<n$ with $\S xs/_j=0$, and 
$\S ys/_j=\S xs/_j$ in all other cases. Since $\S xs/_j$
can be zero when $\S ys/_j$ is not, simulation $F$ does 
not preserve quiescence. For instance, if 
$\S xs/ = (1,0,0,1,0^\omega)$, we need 
$\S ys/ = (1,2,2,1,0^\omega)$. \boks 

\medbreak
Preservation of quiescence does not occur in 
\cite{AbL91}. Its role is played by the stronger concept 
of internal continuity. We therefore have to clarify the 
relationship between these concepts.
Following \cite{AbL91}, we define a visible 
specification $(K,f)$ to be \emph{internally continuous}
iff every infinite initial execution \S xs/ of $K$
with $ f^\omega(\S xs/)\in\S Obs/(K,f)$ is a behaviour 
of $K$. As the next result shows, internal continuity of 
the target specification implies preservation of 
quiescence by every simulation that yields an 
implementation according to Theorem 0.

\nrule \label{intcontused}
Let $(K,f)$ and $(L,g)$ be visible specifications and 
assume that $(L,g)$ is internally continuous. Let 
$F:K\morph L$ be a simulation with $(F;g)\subseteq f$. 
Then $F$ preserves quiescence.  

\proof Let \S xs/ be a behaviour of $K$. We have to 
provide a behaviour \S ys/ of $L$ with 
$(\S xs/,\S ys/)\in F^\omega$ and 
$Q_K(\S xs/) \subseteq Q_L(\S ys/)$. Since $F$ is a 
simulation, we can choose a behaviour \S ys/ of $L$ with 
$(\S xs/,\S ys/)\in F^\omega$. It remains to prove that
$Q_K(\S xs/) \subseteq Q_L(\S ys/)$.

Let $n\in Q_K(\S xs/) $ be given. Write $ \S xn/=E_n(\S xs/)$ 
and $ \S yn/=E_n(\S ys/)$. Then \S yn/ is an infinite 
initial execution of $L$ with $(\S xn/,\S yn/)\in F^\omega$.
Just as in the proof of Theorem 0, the inclusion 
$(F;g)\subseteq f$ implies that $f(x)=g(y)$ for every 
$(x,y)\in F$. It follows that 
$g^\omega(\S yn/) = f^\omega(\S xn/)$. Since 
$n\in Q_K(\S xs/) $, we have 
$ f^\omega(\S xn/)\in \S Obs/(K,f)$. Theorem 0 implies 
that $(K,f)$ implements $(L,g)$. It therefore follows 
that $ g^\omega(\S yn/)\in \S Obs/(L,g)$. Now,
internal continuity of $(L,g)$ implies that \S yn/ is 
a behaviour of $L$, so that $n\in Q_L(\S ys/)$. \boks

\medbreak The following lemma implies that refinement 
mappings and forward and backward simulations all 
preserve quiescence. 

\nrule\label{H2P3strict}
Every flat simulation $F:K\morph L$ preserves quiescence.

\proof Let $\S xs/\in\S Beh/(K)$. Since $F$ is a 
simulation, there exists $\S ys/\in\S Beh/(L)$ with 
$ (\S xs/,\S ys/)\in F^\omega$. It suffices to prove 
that $ Q_K(\S xs/)\subseteq Q_L(\S ys/)$. Let 
$n\in Q_K(\S xs/)$. Write $ \S xn/=E_n(\S xs/)$ and
$ \S yn/=E_n(\S ys/)$. Since 
$n\in Q_K(\S xs/)$, we have $\S xn/\in\S Beh/(K)$.
On the other hand, \S yn/ is an infinite initial 
execution of $L$ and $(\S xn/,\S yn/)\in F^\omega$. 
Flatness of $F$ implies that \S yn/ is a behaviour of 
$L$. This proves $n\in Q_L(\S ys/)$. \boks

\section {An Eternity Variable for Refinement}
\label{sec:refine}

We now develop an alternative for prophecy variables
or backward simulations that is simpler and in a 
theoretical sense more powerful. Extending the metaphor 
of history and prophecy variables, they are named 
eternity variables, since they do not change during 
execution. 

They are simpler than prophecy variables in the sense 
that, below, both the proof of soundness in Lemma 
\ref{eternalsound} and the proof of completeness in 
Theorem 1 are simpler than the corresponding proofs for 
prophecy variables. They are theoretically more powerful 
in the sense that their completeness does not require 
additional finiteness assumptions. 

The idea is that an eternity variable has an 
indeterminate constant value, but that the states 
impose restrictions on this value. A behaviour in 
which the eternity variable ever has a wrong value is 
simply discarded. Therefore, in every behaviour, the 
eternity variable always has a value that satisfies 
all restrictions of the behaviour.

The specification obtained by adding an eternity 
variable is called an eternity extension. In Sect.\ 
\ref{eterdef}, we introduce eternity extensions, prove 
their soundness, and give a simple example. Completeness 
of eternity extension is proved in Sect.\ \ref{eterComp}. 
At first sight, the use of eternity variables may seem to
require arguing about complete behaviours rather than 
states and the next-state relation. As argued in Sect.\ 
\ref{behaveORassert}, however, it is possible to 
combine the use of eternity variables conveniently with 
assertional methods.

\subsection{Eternity Extensions Defined}
\label{eterdef}

Let $K$ be a specification. Let $M$ be a set of 
values for an eternity variable \T m/. 
A binary relation $R$ between $\sigma K$ and $M$ is 
called a \emph{behaviour restriction} of $K$ iff, for 
every behaviour \S xs/ of $K$, there exists 
an $m\in M$ with $(\S xs/_i,m)\in R$ for all indices 
$i\:$:
\begin{tab}
(BR) \> $ \S xs/ \in\S Beh/(K)\IMPLIES
 (\ex m::(\all i:: (\S xs/_i,m)\in R)) $ .
\end{tab}
If $R$ is a behaviour restriction of $K$, we define 
the corresponding \emph{eternity extension} as the 
specification $W$ given by 
\begin{tab}
\>\+ $ \sigma W \IS R $ ,\\
 $ \alpha W \IS R\,\cap(\alpha K\times M)$ ,\\
 $ ((x,m),(x',m'))\in\nu W\EQ 
 (x,x')\in\nu K\LAND m = m' $ ,\\
 $ \S ys/\in\pi W \EQ \S fst/^\omega (\S ys/)\in\pi K
$ .
\end{tab}
It is clear that $\nu W$ is reflexive and that 
$\pi W$ is a property. Therefore $W$ is a 
specification. It is easy to verify that 
$\S fst/:\sigma W\to \sigma K$ is a refinement 
mapping. The soundness of eternity extensions
is expressed by

\nrule\label{eternalsound}
Let $R$ be a behaviour restriction. Then relation 
$\S cvf/=\S cv/(\S fst/)$ is a flat simulation 
$K\morph W$.

\proof We first prove that \S cvf/ is a simulation. 
Let $\S xs/ \in\S Beh/(K)$. We have to construct 
$\S ys/\in\S Beh/(W)$ with 
$(\S xs/,\S ys/)\in \S cvf/^\omega$. 
By (BR), we can choose $m$ with
$(\S xs/_i,m)\in R$ for all $i$.
Then we define $\S ys/_i=(\S xs/_i,m)$. A trivial 
verification shows that the list \S ys/ constructed 
in this way is a behaviour of $W$ with 
$(\S xs/,\S ys/)\in \S cvf/^\omega$. This proves that 
\S cvf/ is a simulation. Flatness of \S cvf/ follows 
directly from the definitions of flatness and $\pi W$. 
\boks

\medbreak The simulation $\S cvf/:K\morph W$ of Lemma
\ref{eternalsound} is called the eternity extension of 
$K$ corresponding to behaviour restriction $R$. 
In this construction, we fully exploit the ability to 
consider specifications that are not machine closed. 
Initial executions of $W$ that cannot be extended to 
behaviours of $W$ are simply discarded. 

\sbreak\S Remark./ If $M$ is a singleton set, such as 
the type \T void/, the existential quantification in 
(BR) can be eliminated and condition (BR) reduces to 
the requirement that $D=\{x\,|\,(x,\_\,)\in R\}$ is an 
invariant. Then $W$ is isomorphic to the $D$-restricted 
specification $K_D$ and \S cvf/ corresponds to the 
simulation $1_D:K\morph K_D$ of Lemma \ref{invars}(a)
in \ref{invarRes}. \boks

\mbreak\S Example J/. We give a simple example where 
a nontrivial eternity variable is used to prove that a 
given relation is a simulation. Let $K$ be the 
specification given by the program 
\begin{tab}
\>\+ $ \B var /\;\T j/:\S Nat/\;,\;
\T b/:\S Boolean/ $ ;\\
$ \B initially: / \T j/=0\Land \neg\,\T b/$ ;\\
$\B do /\;\neg \,\T b/ \TO \T j/:= \T j/+1 $ ;\\
$\bar\quad \T j/ \ne 0 \TO \T b/:=\true $ ;\\
$ \B od/ $ ;\\
\B prop: /\T b/ is sometimes \true.
\end{tab}
Let $L$ be the specification given by
\begin{tab}
\>\+ $ \B var /\;\T k/, \T n/:\S Nat/ := 0, 0 $ ;\\
$\B do /\;\T n/ = 0 \TO \T k/:= 1\;;\;
\S choose /\; \T n/ \geq 1 $ ;\\
$\bar\quad \T k/ < \T n/ \TO \T k/:=\T k/ + 1 $ ;\\
$ \B od/ $ ;\\
\B prop: / sometimes $\T k/=\T n/$.
\end{tab}
Recall that the alternatives in the \B do/ loop are regarded 
as atomic. 
Let relation $F$ between the state spaces of $K$ and $L$ be 
given by
\begin{tab}
\> $ ((j,b),(k,n))\in F \EQ j = k$ .
\end{tab}
We claim that $F$ is a simulation. In every behaviour, 
specification $L$ chooses the number of nontrivial steps of 
the behaviour in the first nontrivial step. For $K$, this 
number is determined in the last nontrivial step. It thus 
needs prescience to construct the behaviour of $L$ from 
that of $K$. 

We therefore factor relation $F$ over an eternity extension. 
For this purpose, we form the eternity extension with 
eternity variable $\T m/:\Nat$ and relation 
\begin{tab}
\> $R:\quad \T j/\leq \T m/\LAND 
(\neg\,\T b/\LOR \T j/=\T m/)$ .
\end{tab}
The state of $K$ remains constant once \T b/ has become 
\true. Therefore, every behaviour of $K$ has a unique 
value for \T m/ that satisfies $R$, namely the final 
value of \T j/. This shows that $R$ is a behaviour 
restriction. We thus form the corresponding eternity 
extension $\S cvf/: K\morph W$. In view of behaviour 
restriction $R$, specification $W$ can be regarded as 
the program 
\begin{tab}
\>\+ $ \B var /\;\T j/, \T m/:\S Nat/\;,\;
\T b/:\S Boolean/ $ ;\\
$ \B initially: / \T j/=0\Land \neg\,\T b/$ ;\\
$\B do /\;\neg \,\T b/\Land \T j/ < \T m/ \TO 
\T j/:= \T j/+1 $ ;\\
$\bar\quad \T j/ = \T m/ \ne 0 \TO \T b/:=\true $ ;\\
$ \B od/ $ ;\\
\B prop: /\T b/ is sometimes \true.
\end{tab}
Let $g:\sigma W \to \sigma L$ be given by 
\begin{tab}
\> $ g(j,b,m) \IS (j, (j=0\,?\;0: m)) $ ,
\end{tab}
where $(\_ \:?\:\_:\_ )$ stands for a conditional 
expression as in the language C. It is easy to see that $g$ 
maps the initial states of $W$ into the initial state of $L$.
Every step according to the first alternative of $W$ is 
transformed into a step of $L$. Every step according to the 
second alternative of $W$ is transformed into a stuttering 
step of $L$. Every behaviour of $W$ is transformed into a 
behaviour of $L$. Therefore $g$ is a refinement mapping 
$W\morph L$. The composition $(\S cvf/;g)$ is contained in 
relation $F$. This shows that $F$ is a simulation. \boks

\subsection{Completeness of Eternity Extensions}
\label{eterComp}

The combination of forward simulations, 
eternity extensions and refinement mappings is 
semantically complete in the following sense. 

\bbreak \B Theorem 1./ Let $F:K\morph L$ be a simulation
that preserves quiescence. There exist a forward simulation
$\S fw/:K\morph H$, an eternity extension $\S et/:H\morph W$ 
and a refinement mapping $ g:W\morph L$ such that 
$ (\S fw/;\S et/;g)\subseteq F$.

\proof According to Lemma \ref{unfold}, the unfolding 
$\S cvl/: K\morph K^\#$ is a forward simulation. It 
therefore suffices to prove the following more specific 
result.   

\nrule\label{thmUnfold} Let $F:K\morph L$ be a simulation 
that preserves quiescence. The unfolding 
$\S cvl/:K\morph K^\#$ has an eternity extension 
$\S cvf/:K^\# \morph W$ and a refinement mapping 
$ g:W\morph L$ such that 
$ (\S cvl/;\S cvf/;g)\subseteq F$.

\proof We extend $K^\#$ with an eternity variable \T m/ 
in the set $ \S Beh/(L) $. For this purpose, let relation 
$R$ between $\sigma {K^\#}$ and $\S Beh/(L)$ consist of
the pairs $(\S xs/,\S ys/)$ such that, for some 
$\S xt/\in\S Beh/(K)$, it holds that
\begin{tab}
\> $ \S xs/ \sqleq \S xt/ \Land 
 (\S xt/,\S ys/)\in F^\omega \Land 
 Q_K(\S xt/)\subseteq Q_L(\S ys/) $ .
\end{tab}
We show that $R$ is a behaviour restriction by 
verifying condition (BR). 
Let \S uss/ be any behaviour of 
$K^\#$. Define \S vss/ to be the stutterfree behaviour of 
$K^\#$ with $\S vss/ \preceq \S uss/$. 
By Lemma \ref{bloweddown}, we have that 
$ \S xt/ = \S last/^\omega(\S vss/)$ is a behaviour of 
$K$ such that $\S vss/_i$ is a prefix of \S xt/ for all 
indices $i$. Since $F:K\morph L$ preserves quiescence, 
specification $L$ has a behaviour \S ys/ 
with $(\S xt/,\S ys/)\in F^\omega$ and 
$Q_K(\S xt/)\subseteq Q_L(\S ys/)$. 
This implies that $(\S vss/_i,\S ys/) \in R$ for all 
$i\in\Nat$. Since every element of \S uss/ is an element 
of \S vss/, it follows that $(\S uss/_i,\S ys/) \in R$ 
for all $i\in\Nat$. Taking $\T m/=\S ys/$, this proves 
condition (BR), so that $R$ is a behaviour restriction.  

Let $W$ be the $R$-eternity extension of $K^\#$. By 
Lemma \ref{eternalsound}, we have a simulation 
$\S cvf/:K^\#\morph W$. Define $g:R\to \sigma L$ by 
\begin{tab}
\>$g(\S xs/,\S ys/) = \S last/(\S ys/\,|\,\ell(\S xs/))$ .
\end{tab}
We show that $g$ is a refinement mapping from $W$ to 
$L$. Firstly, let $w\in\alpha W$. Then $w$ is of the form 
$w = (\S xs/,\S ys/) $ with $\ell(\S xs/) = 1$. Therefore
$g(w) = \S last/(\S ys/|1) = \S ys/_0\in\alpha L$. In 
every nonstuttering step in $W$, the length of \S xs/ is 
incremented with 1 and then we have 
$(\S ys/_n,\S ys/_{n+1})\in\nu L$. Therefore, function 
$g$ maps steps of $W$ to steps of $L$. 

In order to show that $g$ maps every behaviour of $W$ to 
a behaviour of $L$, it suffices to show that 
$g^\omega(\S ws/)\in\pi L$ for every stutterfree 
behaviour of $W$. So, let \S ws/ be a stutterfree 
behaviour of $W$. 
Since \S ws/ is a behaviour of $W$, its elements have a 
common second component $\S ys/\in\S Beh/(L)$. We can 
therefore write $\S ws/_k=(\S us/_k,\S ys/)$ for all $k$.
Since $\S ws/\in\S Beh/(W)$, we have 
$\S us/ = \S fst/^\omega(\S ws/)\in \S Beh/(K^\#)$. 
In particular, $(\S us/_k,\S us/_{k+1})\in\nu {K^\#}$ 
for all $k$, and $\S last/^\omega(\S us/)\in\pi K$. 

We have $g(\S ws/_k) = \S last/(\S ys/|\ell(\S us/_k))$.
Since \S ws/ is stutterfree, \S us/ is stutterfree. 
There are two possibilities. Either all elements of \S us/ 
are different or up to some index $n$ all elements of \S us/ 
are different and from $n$ onward they stay the same. This
implies, that either $\ell(\S us/_k)=k+1$ 
for all $k$, or there exist a number $n$, such 
that $ \ell(\S us/_k) = \R min/(k,n)+1$ for all $k$. 
In the first case, we have 
$g^\omega(\S ws/)= \S ys/\in \pi L $. In the second 
case, $g^\omega(\S ws/) = E_n(\S ys/)$. Therefore, 
$g^\omega(\S ws/)\in\pi L$ would follow from 
$n\in Q_L(\S ys/)$. Since 
$(\S us/_n,\S ys/) = \S ws/_{n}\in R$, there 
exists a behaviour \S ut/ of $K$ such that 
$\S us/_n \sqleq \S ut/$ and $(\S ut/,\S ys/)\in F^\omega$ 
and $Q_K(\S ut/)\subseteq Q_L(\S ys/)$. Since 
$ \ell(\S us/_n) = n+1$, we have $\S us/_{n}=(\S ut/|n+1)$. 
This implies that $E_n(\S ut/)$ equals $\S us/_n$ 
followed by infinitely 
many states $\S ut/_n=\S last/(\S us/_n)$. It follows that
$E_n(\S ut/) = \S last/^\omega(\S us/)\in\pi K$ 
and hence $n\in Q_K(\S ut/)\subseteq Q_L(\S ys/)$.

It remains to prove $(\S cvl/;\S cvf/;g) \subseteq F$.
Let $(x,y)$ be in the lefthand relation. By the 
definition of $(\S cvl/;\S cvf/;g)$, there 
exist $\S xs/\in\sigma {K^\#}$ and $w\in\sigma W$ with 
$x = \S last/(\S xs/)$ and $\S xs/=\S fst/(w)$ and 
$g(w)= y$. By the definition of $W$, we can choose 
$\S ys/\in\S Beh/(L)$ with $w=(\S xs/, \S ys/)$. Let 
$n=\ell(\S xs/)$. Then $x = \S xs/_{n-1}$ and 
$y=g(w)=\S ys/_{n-1} $. Since $(\S xs/,\S ys/)\in R$, we 
also have $(x,y)= (\S xs/_{n-1}, \S ys/_{n-1})\in F$. 
This proves the inclusion. \boks

\mbreak\S Remarks./ Theorem 1 is more relevant than Lemma 
\ref{thmUnfold} since it suggests the flexibility to add 
conveniently many history variables, and not more than 
necessary. 

The converse of Theorem 1 also holds. In fact, forward 
simulations, eternity extensions and refinement mappings 
are flat simulations, which preserve quiescence by Lemma 
\ref{H2P3strict}. Since preservation of quiescence is 
compositional, it follows that every simulation $F$ that 
satisfies the consequent of Theorem 1 preserves quiescence.

\subsection {Behavioural or Assertional Reasoning?}
\label{behaveORassert}

In general, there are two methods for the verification of 
concurrent algorithms (as discussed, e.g., in \cite{Hes02a} 
p.\ 344). One method, the assertional approach, is to rely 
on invariants and variant functions. The alternative, the 
behavioural approach, is to argue about execution sequences 
(behaviours) where certain actions precede other actions. 
We prefer the assertional approach, see also \cite{H98b} 
where we described it as the synchronic approach. Yet, it 
is clear that, in the analysis of an algorithm that 
gradually modifies the state, we cannot avoid temporal 
or behavioural arguments completely. We therefore strive at 
a separation of concerns where the behavioral argument is 
a formal triviality and all complexity of the algorithm is 
treated at the level of states and the next-state relation. 

One may object that our proof obligation (BR) in \ref 
{eterdef} requires quantification over all possible 
behaviours, which is precisely what the assertional 
methods try to avoid. 
This objection is not justified. In fact, it could equally 
well be raised against the use of invariants, defined as 
predicates that hold in all reachable states. 

The question thus boils down to establishing condition (BR) 
of \ref{eterdef}. Given a behaviour \S xs/, one has to 
construct a value $m$ for the eternity variable such that 
$(\all i:: (\S xs/_i,m)\in R)$. 
In practice, we proceed as follows. First rephrase 
$(\all i:: (\S xs/_i,m)\in R)$ as 
$(\all i :: \S xs/_i \in R(m))$ for a state predicate 
$R(m)$, with a free variable $m$ yet to be determined. 
Predicate $R(m)$ plays the same role as an invariant, 
but only for a specific behaviour \S xs/. 

We now use that Theorem 1 allows us the introduction of 
history variables. We introduce a history variable the 
value of which converges in a certain sense for every 
behaviour, and we use the ``limit'' as a value for $m$. 
In the above example I, the final value of the variable 
\T j/ was this limit. 

In our more interesting examples (see Sect.\ \ref{example} 
and \cite{Hes03}), the eternity variable $m$ is an infinite 
sequence and the approximating history variable consists of 
a pair $(\T n/, \T a/)$ where \T n/ holds a natural number 
and \T a/ is an infinite array filled upto \T n/. 
This pair is modified only by steps of the form
\begin{tab}
\> $\la\; \T a/[\T n/]:= \S expression/\;;\;
\T n/:= \T n/+1\;\ra$ .
\end{tab}
The behaviour restriction is given as the state predicate
\begin{tab}
(*) \> $R(m) \EQ (\all j: j < \T n/: m(j) = \T a/[j])$ .
\end{tab}
Since \T n/ is incremented only and \T a/ is never modified 
at indices below \T n/, for every behaviour \S xs/, the 
existence of a value $m$ that always satisfies $R(m)$ is a 
formal triviality. 

In our applications, this is the only behavioural argument 
needed. The remainder of the verification can be done by 
assertional methods. Of course, creativity is needed to 
come up with approximating history variables that carry 
enough information, but this is the same kind of creativity 
as needed to invent invariants. 

When we restrict the method to behaviour restrictions of the 
special kind (*), we cannot maintain completeness, since in 
the proof of Lemma \ref{thmUnfold} we used a different kind 
of behaviour restriction. So, indeed, we cannot guarantee 
that in all applications there is a convenient reduction to 
the assertional setting.

\section{A Slightly Bigger Example}
\label{example}

In this section, we illustrate the theory by a tiny 
application. We prove that a relation between the state 
spaces of specifications \S K0/ and \S K1/ is a simulation 
by factoring it over the forward simulation, an eternity 
extension, an invariant restriction, and two refinement 
mappings. 

\subsection{The Problem}

Let \S K0/ be the specification corresponding to the 
program 
\begin{tab}
\>\+ $ \B var /\;\T j/:\S Nat/:= 0 $ ;\\
$\B do /\;\true \TO \T j/ :=\T j/ + 1$ ;\\
$\;\bar\quad \T j/ >0\TO \T j/ := 0 $ ;\\
$\B od/ $ ;\\
\B prop: /\T j/ decreases infinitely often.
\end{tab}
The fairness assumption requires that the second 
alternative is chosen infinitely often. Specification 
\S K0/ has $\sigma {\S K0/}=\Nat$ and 
$\alpha{\S K0/} = \{0\}$ and relation $\nu {\S K0/}$ 
given by 
\begin{tab}
\> $ (j,j')\in \nu {\S K0/} \EQ j'\in\{0,j,j+1\}$ .
\end{tab}
The supplementary property that \T j/ decreases 
infinitely often, is expressed in 
$\pi{\S K0/} = \Box\Diamond\sem > $.

We extend specification \S K0/ with a variable \T z/ that
guesses when \T j/ will jump back. We thus obtain the 
extended specification \S K1/ with the program 
\newpage\begin{tab}
\>\+ $ \B var /\;\T j/, \T z/:\S Nat/:= 0, 0 $ ;\\
$\B do /\;\T j/<\T z/ \TO \T j/ :=\T j/ + 1$ ;\\
$\;\bar\quad \T j/ = 0\TO 
\T j/ := 1\;;\; \S choose /\;\T z/ \geq 1 $ ;\\
$\;\bar\quad \T j/ = \T z/\TO 
\T j/ := 0\;;\; \T z/ := 0 $ ;\\
$\B od/ $ ;\\
\B prop: /$(\T j/,\T z/)$ changes infinitely often. 
\end{tab}
Recall that the alternatives in the loop are executed 
atomically. The supplementary property only ensures that 
behaviours do not stutter indefinitely. 
We thus have $\pi {\S K1/}=\Box\Diamond\sem \ne$.

The function $f_{1,0}:\sigma {\S K1/}\to \sigma{\S K0/}$ 
given by $f_{1,0}(j,z)=j$ is easily seen to be a 
refinement mapping $\S K1/\morph \S K0/$. 

More interesting is the converse relation 
$F_{0,1}=\S cv/(f_{1,0})$. It is not difficult to show 
by ad-hoc methods that $F_{0,1}$ is a simulation 
$\S K0/\morph\S K1/$, but the aim of this section is to 
do it systematically by means of the theory developed.

In comparison with \S K0/, the variable \T z/ seems to 
prophecy the future behaviour. This suggests to use a 
backward simulation. Our best guess is the relation 
$F=\{(j, (j,z))\mid j\leq z\}$ between the state spaces 
of \S K0/ and \S K1/. Indeed, relation $F$ satisfies 
three of the four conditions for backward simulations, 
but condition (B2) fails: the sets $\{y\mid(x,y)\in F\}$ 
are always infinite. We therefore use factorization over 
an eternity extension.

\subsection{A History Extension to Approximate Eternity}
\label{exforward}

Every behaviour of \S K1/ contains infinitely many steps 
where a new value for \T z/ is chosen. These values are 
prophecies with respect to \S K0/. In the behaviours of 
\S K0/, these values can only be seen at the jumping steps.
We therefore extend \S K0/ with an infinite array of history 
variables to record the subsequent jumping values. 

We thus extend specification \S K0/ with two history 
variables \T n/ and \T q/. Variable \T n/ counts the 
number of backjumps of \T j/, while \T q/ is an array 
that records the values from where \T j/ jumped.
\begin{tab}
\>\+ $ \B var /\;\T j/:\S Nat/:= 0\;,\;
\T n/:\S Nat/:= 0 $ ,\\
\> $ \T q/:\B array /\S Nat/\B\ of /\S Nat/:=
([\S Nat/]\;0) $ ;\\
$\B do /\;\true \TO \T j/ :=\T j/ + 1 $ ;\\
$\;\bar\quad \T j/ >0\TO \T q/[\T n/] := \T j/\;;\;
\T n/:=\T n/ + 1\;;\; \T j/ := 0 $ ;\\
$\B od/ $ ;\\
\B prop: /\T j/ decreases infinitely often.
\end{tab}
This yields a specification \S K2/ with the 
supplementary property $\Box\Diamond\sem{\T j/>\T j/'}$
where $\T j/'$ stands for the value of \T j/ in the 
next state. 

It is easy to verify that the function 
$f_{2,0}: \sigma{\S K2/}\to\sigma{\S K0/}$ given by 
$f_{2,0}(j,n,q) = j$ is a refinement mapping. Its 
converse $F_{0,2} = \S cv/(f_{2,0})$ is a forward 
simulation $\S K0/\morph \S K2/$. Indeed, the conditions 
(F0) and (F2) hold almost trivially. As for (F1), if we 
have related states in \S K0/ and \S K2/, and the state 
in \S K0/ makes a step, it is clear that \S K2/ can take 
a step such that the states remain related. The 
variables \T n/ and \T q/ are called history variables 
since they record the history of the execution. 

\subsection{An Example of an Eternity Extension}
\label{exetern}

We now extend \S K2/ with an eternity variable \T m/, 
which is an infinite array of natural numbers with the 
behaviour restriction 
\begin{tab}
\> $R:\quad (\all i: 0\leq i < \T n/: \T m/[i] = \T q/[i])$ .
\end{tab}
We have to verify that every behaviour of \S K2/ 
allows a value for \T m/ that satisfies condition 
$R$. So, let \S xs/ be an arbitrary behaviour of 
\S K2/. Since \T j/ jumps back infinitely often in 
\S xs/, the value of \T n/ tends to infinity. This 
implies that $\T q/[i]$ is eventually constant for 
every index $i$. We can therefore define function 
$m:\Nat\to\Nat$ by $\Diamond\Box\sem{m(i)=\T q/[i]}$
for all $i\in\Nat$. It follows that 
$\Box\sem{i<\T n/\:\implies\: m(i)=\T q/[i]}$
for all $i$. This proves that $m$ is a value for \T m/ 
that satisfies $R$ for behaviour \S xs/. 

Let \S K3/ be the resulting eternity extension and 
$F_{2,3}:\S K2/\morph\S K3/$ be the simulation induced 
by Lemma \ref{eternalsound}. Specification \S K3/ 
corresponds to the program 
\begin{tab}
\>\+ $ \B var /\;\T j/:\S Nat/:= 0\;,\;
\T n/:\S Nat/:= 0 $ ,\\
\> $ \T q/:\B array /\S Nat/\B\ of /\S Nat/:=
([\S Nat/]\;0) $ ,\\
\> $ \T m/:\B array /\S Nat/\B\ of /\S Nat/\;
\{\S arbitrary/\} $ ;\\
$\B do /\;\true \TO \T j/ :=\T j/ + 1 $ ;\\
$\;\bar\quad \T j/ = \T m/[\T n/] >0 
\TO \T q/[\T n/] := \T j/\;;\;
\T n/:=\T n/ + 1\;;\; \T j/ := 0 $ ;\\
$\B od/ $ ;\\
\B prop: /\T j/ decreases infinitely often.
\end{tab}

\subsection {Using Refinement Mappings and an Invariant}

We first eliminate array \T q/, which has played its role.
This gives a refinement mapping $f_{3,4}$ from \S K3/ to 
the specification \S K4/ with program 
\begin{tab}
\>\+ $ \B var /\;\T j/:\S Nat/:= 0\;,\;
\T n/:\S Nat/:= 0 $ ,\\
\> $ \T m/:\B array /\S Nat/\B\ of /\S Nat/\;
\{\S arbitrary/\} $ ;\\
$\B do /\;\true \TO \T j/ :=\T j/ + 1 $ ;\\
$\;\bar\quad \T j/ = \T m/[\T n/] >0 
\TO \T n/:=\T n/ + 1\;;\; \T j/ := 0 $ ;\\
$\B od/ $ ;\\
\B prop: /\T j/ decreases infinitely often.
\end{tab}
Since \T j/ must decrease infinitely often in \S K4/, the 
occurring states of \S K4/ satisfy the invariant 
\begin{tab}
\> $D:\quad \T j/\leq \T m/[\T n/]\Land 
(\all i:: \T m/[i]\geq 1) $ .
\end{tab}
Note that $D$ is not a forward invariant of \S K4/, see 
Sect.\ \ref{specs}. Let \S K5/ be the $D$-restriction of 
\S K4/, with the simulation $1_D:\S K4/\morph \S K5/$ of  
Lemma \ref{invars}(a). Specification \S K5/ corresponds to 
\begin{tab}
\>\+ $ \B var /\;\T j/:\S Nat/:= 0\;,\;
\T n/:\S Nat/:= 0 $ ,\\
\> $ \T m/:\B array /\S Nat/\B\ of /\S Nat/\S\ with /
(\all i:: \T m/[i]\geq 1)  $ ;\\
$\B do /\;\T j/ < \T m/[\T n/] \TO \T j/ :=\T j/ + 1 $ ;\\
$\;\bar\quad \T j/ = \T m/[\T n/] >0 
\TO \T n/:=\T n/ + 1\;;\; \T j/ := 0 $ ;\\
$\B od/ $ ;\\
\B prop: /\T j/ decreases infinitely often.
\end{tab}
Let function $f_{5,1}: \sigma{\S K5/}\to\sigma{\S K1/}$ 
be defined by 
\begin{tab}
\> $ f_{5,1}(j,n,m) \IS (j,(j=0\,?\;0:m[n])) $ ,
\end{tab}
again using a C-like conditional expression. We verify 
that $f_{5,1}$ is a refinement mapping.
Since $f_{5,1}(0,0,m)=(0,0)$, initial states are mapped 
to initial states.
We now show that a step of \S K5/ is mapped to a step 
of \S K1/. By convention, this holds for a stuttering 
step. A nonstuttering step that starts with $\T j/=0$ 
increments \T j/ to 1. The $f_{5,1}$-images make a 
step from $(0,0)$ to $(1,r)$ for some positive number $z$. 
This is in accordance with \S K1/. A step of \S K5/ 
that increments a positive \T j/ has the precondition 
$\T j/<\T m/[\T n/]$; therefore, the 
$f_{5,1}$-images make a \S K1/-step. A back-jumping 
step of \S K5/ has precondition $\T j/ = \T m/[\T n/]>0$.
Again, the $f_{5,1}$-images make a \S K1/-step. It 
is easy to see that $f_{5,1}$ transforms behaviours of 
\S K5/ to behaviours of \S K1/.

We thus have a composed simulation 
$G=(F_{0,2};F_{2,3};f_{3,4};1_D;f_{5,1}):\S K0/\morph\S K1/$.
One can verify that $(j,(k,m))\in G$ implies $j=k$. 
It follows that the above relation $F_{0,1}$ satisfies 
$G \subseteq F_{0,1}$. Therefore, $F_{0,1}$ is a simulation 
$\S K0/\morph \S K1/$. This shows that an eternity 
extension can be used to prove that $F_{0,1}$ is a 
simulation $\S K0/\morph\S K1/$.

\mbreak\S Remark./ We have taken more steps here than 
accounted for in Theorem 1. By taking a different behaviour
restriction $R$, we could have compressed the last three 
steps into one more complicated step. \boks

\section{Conclusions and Future Work} \label{conclusion}

We have introduced simulations of specifications to 
unify all cases where an implementation relation can be 
established. This unifies refinement mappings, history 
variables or forward simulations, and prophecy variables 
or backward simulations, and refinement of atomicity as 
in Lipton's Theorem \cite{CoL98,Lip75}. This unification 
is no great accomplishment: a general term to unify 
distinct kinds of extensions is useful for the 
understanding, but methodologically void.  

We have introduced eternity extensions as variations of 
prophecy variables and backward simulations. We have 
proved semantic completeness: every simulation that 
preserves quiescence can be factored as a composition 
of a forward simulation, an eternity extension and a 
refinement mapping. The restrictive assumptions 
machine-closedness and finite invisible nondeterminism, 
as needed for completeness of prophecy variables or 
forward-backward simulations in \cite{AbL91,LyV95} are 
superfluous when eternity variables are allowed. The 
assumption of internal continuity is weakened to 
preservation of quiescence. 

The theory has two versions. In the strict version 
presented here, we allow the concrete behaviours to take 
more but not less computation steps than the abstract 
behaviours. This is done by allowing additional 
stutterings to the abstract specifications. The strict 
theory is also the simpler one and it results in a finer 
hierarchy of specifications than the stuttering theory.

It is likely that the results of this paper can be 
transferred to input-output automata and labelled 
transition systems. The ideas may also be useful in 
specifications and correctness arguments for real-time 
systems. 

As indicated above, we developed the theory of eternity 
variables to apply them in \cite{Hes03} to the serializable 
database interface problem of \cite{Bro92,Lam92,Sch92}.
The practicality of the use of eternity variables is 
witnessed by the fact that the proof in \cite{Hes03} 
is verified by means of the mechanical theorem prover 
{\small NQTHM} \cite{BM97}, which is based on first-order 
logic. 

\sbreak {\small{\bf Acknowledgements.} I am grateful to 
Eerke Boiten,  Leslie Lamport, Caroll Morgan, and Gerard 
Renardel de Lavalette for encouragements, comments, and 
profound discussions.}

\bbreak
Received August 2002;
revised August 2003;
accepted August 2003

\end{document}